\documentclass{article}

\usepackage{setspace} 
\usepackage{appendix}
\usepackage{amsmath}
\usepackage{amsfonts} 
\usepackage{amsthm}
\usepackage{graphicx}
\usepackage[colorlinks=true, allcolors=blue]{hyperref}

\usepackage[english]{babel}

\usepackage[letterpaper,top=2cm,bottom=2cm,left=3cm,right=3cm,marginparwidth=1.75cm]{geometry}
\newtheorem{theorem}{Theorem}[section]

\newtheorem{lemma}[theorem]{Lemma}
\newtheorem{axiom}{Axiom}[section]

\usepackage[bibstyle=numeric, 
            citestyle=authoryear, 
            backend=biber, 
            natbib=true]
            {biblatex}

\title{Expected Growth Criterion: An Axiomatization}
\author{Joshua Lawson}
\begin{document}
\maketitle

\begin{abstract}
    I provide necessary and sufficient conditions for an agent's preferences to be represented by a unique ergodic transformation. Put differently, if an agent seeks to maximize the time average growth of their wealth, what axioms must their preferences obey? By answering this, I provide economic theorists a clear view of where "Ergodicity Economics" deviates from established models.
\end{abstract}

{\bf JEL CLassification:} C02, C46, D81, G00, G11


\section{Introduction}
    In \citet{mederErgodicitybreakingRevealsTime2021}, neurobiologist Oliver Hulme and his colleagues reported strong evidence that agents dynamically adapt their utility functions such that changes in wealth are rendered ergodic. When facing a multiplicative dynamic process, the subjects' implied utility functions were approximately logarithmic. When faced with an additive process, their utility equalled their wealth. In the language of economics, it was as if the agents suddenly changed from being risk averse to risk neutral.\footnote{Appendix \ref{appendix:CE} provides a very succinct description of the experiment. Ideally, one would be more familiar with the experiment before proceeding.} Proponents of a new entrant to decision theory, Ergodicity Economics (EE), claimed these results as proof positive that agents prefer to maximize the time-average growth rate of their wealth and that the predictions of their null model far outperform those of classical expected utility theory (EUT). This assertion has been met with strong criticism. Skeptics have argued that the predictions of EE have been unfairly compared to those of static EUT. Notably, \citet{doctorEconomistsViewsErgodicity2020} and \citet{doctorEconomistsViewsErgodicity2020a} posit that the results should be compared normatively to those of multi-period expected utility and descriptively to those of dynamic ambiguity aversion. This raises two fundamental questions, (1) axiomatically, what does EE require and (2) to which theory of decision making should it be compared?  Therefore, my primary contribution is to axiomatize ergodicity economics as described in \citet{petersTimeInterpretationExpected2021} and applied in the Copenhagen Experiment (CE), thus supplying terra firma to future criticisms. Additionally, appendix \ref{appendix:comp_models} takes the advice of \citet{doctorEconomistsViewsErgodicity2020} and compares the foregoing experiment to multi-period EUT.

    \subsection{Ergodicity Economics}
    
        Starting in 2012, German Physicist Ole Peters and his colleagues at the London Mathematical Laboratory (LML) began publishing a collection of articles with an ambitious goal: "a re-write of economic theory."\footnote{\hyperlink{0}{https://ergodicityeconomics.com/}} Their project obtains its moniker from the center piece of its purported insights, ergodicity - a characteristic of dynamic processes they claim is untreated in the history of decision theory. In \citet{petersErgodicityProblemEconomics2019}, the author defines an ergodic process $f$ as one where 

            \begin{equation}
                \label{equation:ergodicity}
                \lim_{t \to \infty} \frac{1}{T} \int_0^Tf(t)dt = \int_{\Omega}f(\omega)d\mu,
            \end{equation}

        i.e., a process where the time average converges to it's expected value. Beginning with this definition and a seemingly innocuous assumption that agents seek to maximize the time average growth rate of their wealth, they argue that expected utility theory as a whole is nonsensical and therefore so is anything built on top of it. Unsurprisingly, this has been met with incredulity from most economists. While most of the arguments in \citet{petersErgodicityProblemEconomics2019} are of a philosophical nature, a set of more technical arguments can be found in \citet{petersTimeInterpretationExpected2021}. In that article, the authors assume an agent's wealth evolves temporally according to an exogenously defined stochastic process and that agents have the same objective: maximize the time average growth rate of their wealth. They go on to show how one can map between utility functions and stochastic processes.

        \citet{doctorEconomistsViewsErgodicity2020} and its supplementary appendix retort that at best, the claims by Peters and company are well known across decision theory and that at worst, they stem from confusion about what EUT does and does not require. A staunch advocate of EE might quip that their paradigm makes any extension of utility theory beyond the 18th century unnecessary. Indeed, this is even alluded to in \citet{petersTimeInterpretationExpected2021}. Alternatively, one could take the position that they wish to minimize arguments that rely heavily on psychology and stretch the boundaries of what can be done with observable data. And yet others will be curious how after we have shifted to this new paradigm, how can we reincorporate the notions of risk aversion, impatience, and other psychological factors to obtain sharper predictions? This new framework offers tantalizing opportunities, a full exposition of which is beyond the scope of this manuscript.

\section{Mathematical Preliminaries}

    \subsection{Preference Relations}
        Given a nonempty set $X$ with elements $x,y$, let $\succeq$ be a binary relation on $X$ representing the statement "is at least as preferred as". Let $\sim$ denote indifference, that is, the symmetric part of $\succeq$, where $x \sim y$ if $x \succeq y$ and $y \succeq x$. I will say $\succeq$ is degenerate if $x \sim y \forall x, y \in X.$ I will call $\succeq$ a complete preorder if it satisfies the usual properties of reflexivity, transitivity, and completeness. For any $x, y, z \in X:$
            \begin{enumerate}
                \item \textit{Reflexive}: $x \succeq x$
                \item  \textit{Transitive}: $x \succeq y$ and $y \succeq z$ $\implies x \succeq z.$
                \item \textit{Complete:} Either $x \succeq y$, $y \succeq x,$ or both.
            \end{enumerate}

        Additionally, I will say a function $u$ cardinally represents $\succeq$ if it is unique up to a positive affine transformation and $x \succeq y \iff u(x) \geq u(y) \forall x, y \in X.$

    \subsection{Stochastic Wealth Processes}

        In this paper I will suppose time is infinite but discrete, $t \in \mathcal{T}:=[0, 1, ...]$ and I assume the state space $\Omega$ admits a filtration $\mathcal{F}:=\{F_t\}_t,$ where $F_0:=\{ \Omega, \emptyset\}$ and $F_0 \subset F_1 \subset ... \subset F.$ Let us denote the agent's initial endowment as $x_0 \in X \subseteq \mathbb{R}$ and suppose her wealth at time $t$ is an $F_t$ measurable random variable $x_t:\Omega \to \mathbb{R}.$ Then define the $\mathcal{F}$-adapted stochastic wealth process (SWP), $x:=\{x_0, x_1, ...\} = \{x_t\}_t$ and let $\mathcal{X}$ denote the space of all stochastic wealth processes. Finally, I suppose each agent has preferences $\succeq$ defined on $\mathcal{X}$.

    \subsection{Growth Rates}
        
        For an SWP, define the rate of change,

            \begin{equation}
                \label{eq:random_rate}
                \tilde{r}(t):= \frac{x_t - x_0}{t}
            \end{equation}

        Define the sample average rate of change,

            \begin{equation}
                \label{eq:sample_rate}
                \hat{r}(t,N):= \frac{1}{N}\sum\limits_{i=1}^{N}\frac{x_t(\omega_i) - x_0}{t},
            \end{equation}
            
        where $x_t(\omega_i)$ denotes a particular realization of the random variable $x_t$. Additionally, define the time-average rate of change, 

            \begin{equation}
                \label{eq:time_rate}
                \bar{r}:= \lim\limits_{t \to \infty}\hat{r}(t,N)
            \end{equation}

        and the expected rate of change, 

            \begin{equation}
                \label{eq:expected_rate}
                E[r(t)] := \lim\limits_{N \to \infty} \hat{r}(t,N).
            \end{equation}

        Neither limit in the preceding equations is guaranteed to exist, however, for the moment we will pretend they do. As a specific example to illustrate both ergodicity and our definition of growth rates, suppose $x_t = e^{gt + \sigma \epsilon}$ where $g, \sigma \in \mathbb{R}$ and $\epsilon \sim N(0,1).$ Then $\tilde{r} = \frac{e^{gt + \sigma \epsilon} - x_0}{t}$ and 
        
            \begin{equation}
                E[r(t)] = \frac{1}{t}\lim\limits_{N \to \infty}\frac{1}{N}\sum\limits_{i=1}^{N}e^{gt + \sigma \epsilon} - x_0 
            \end{equation}

            \begin{equation}
                = \frac{1}{t}E[e^{gt + \sigma \epsilon} - x_0] 
            \end{equation}

            \begin{equation}
                = \frac{1}{t}[e^{gt + \frac{\sigma^2}{2}} - x_0]
            \end{equation}

        Furthermore, 
        
            \begin{equation}
                \bar{r} = \lim\limits_{t\to \infty} \frac{e^{gt + \frac{\sigma^2}{2}} - x_0}{t} \to \infty
            \end{equation}
            
        Therefore $\bar{r} \neq E[r(t)]$ and the transformation $g(x_t):=x_t$ is not ergodic. In contrast, suppose $f(x_t):=ln(x_t).$ Then $f(x_t) = gt + \sigma \epsilon \implies \tilde{r} = gt + \sigma \epsilon - ln(x_0)$ and
            
            \begin{equation}
                E[\tilde{r}] = \frac{1}{t}E[gt + \sigma \epsilon - ln(x_0)]
            \end{equation}
            
            \begin{equation}
                =g
            \end{equation}

        Moreover, 
        
            \begin{equation}
                \bar{r} = \lim\limits_{t \to \infty}\frac{gt + \sigma \epsilon - ln(x_0)}{t} 
            \end{equation}
         
            \begin{equation}
                = g
            \end{equation}

            \begin{equation}
                \implies  \bar{r} = E[r(t)]. 
            \end{equation}

        Thus, the function $f$ is an ergodic transformation of the underlying process. If $f(x_t) = ln(x_t)$ then maximizing either the time average or expected rate of change will produce the same optimal strategy. One can think of an ergodic transformation as the special function that isolates the coefficient of time for a certain stochastic process. If you remove the concept of ergodicity, the previous example is well known and spawned a contentious literature. In \citet{markowitzInvestmentLongRun1976}, the author suggested that portfolio analysts not present to investors the portion of the mean-variance efficient frontier that lay above the approximate maximum geometric growth rate because these portfolios have greater short-term volatility and less long-run return. At the time, many used log-wealth maximization to claim that only the log transformation was a rational utility function. This argument was a variant of Samuelson's "fallacy of large numbers" (\citet{samuelsonRiskUncertaintyFallacy1963}) and was strongly refuted in \citet{samuelsonFallacyMaximizingGeometric1971} and \citet{samuelsonWhyWeShould1979}. Some interpreted Samuelson's fallacy to mean that if an agent's utility function for a single period game was not logarithmic then it could not be for a repeated game, e.g. \citet{prattProperRiskAversion1987} and \citet{kimballStandardRiskAversion1993}. \citet{rossAddingRisksSamuelson1999} then showed how rejecting a single bet but accepting a sequence of bets can be perfectly rational in EUT, while \citet{benartziRiskAversionMyopia1999} argued that the irrationality of Samuelson's colleague is not attributable to a fallacious use of the law of large numbers but myopic loss aversion and explainable using the prospect theory of Kahneman and Tversky. Regardless, log-wealth maximization became synonymous with growth optimal. \citet{longNumerairePortfolio1990} showed how in the absence of arbitrage, the growth optimal portfolio (GOP) has the special property of being the numeraire that turns the prices of risky assets into martingales with respect to historical probability measures. Attempting to reconcile the works of Kelly and Samuelson, \citet{carrGeneralizedCompoundingGrowth2020} generalized growth optimality beyond maximizing expected log of wealth. They showed that under suitable dynamics, maximizing the power and square-root utility functions can also correspond to growth optimality. However, until Peters and his colleagues, no one had pointed out the key feature that all of these transformations share - they are ergodic transformations of the underlying process. 

\section{Axiomatization}
    
     As a reminder, agents have preferences denoted by $\succeq$ and defined on $\mathcal{X},$ the space of all SWPs. Although the agent is considering dynamic processes, I am not assuming she is making multiple decisions. This is in contrast to the classical representations of Koopmans, Samuelson (discounted utility), and Kreps \& Porteus (temporal vNM) where agents have preferences for intertemporal consumption and timing of uncertainty resolution. An outline of the road ahead is the following: I assume that all uncertainty is objective and resolves at an unknown point in the future. Placing minimal restrictions on the agents preference relation, I arrive at a representation similar to vNM. Finally, I impose what I call the ergodicity axiom (EA), which delivers the desired result. There is almost certainly a relationship between the EA and the temporal consistency axiom of \citet{krepsTemporalResolutionUncertainty1978}, which should be explored in the future.

    The first axiom I impose on preferences is a standard consistency axiom,

        \begin{axiom}[Consistency]
            \label{axiom:consistency}
            \(\succeq\) satisfies completeness, transitivity, and reflexivity on \(\mathcal{X}\).
        \end{axiom}

    Next, define the mixture operator $h:\mathcal{X} \times \mathcal{X} \times (0,1) \to \mathcal{X}$ as $h(x, x';\alpha):= \{(\alpha x_t + (1-\alpha)x_t')_t\}_t.$ The next axiom is critical to ensure continuity of our functional. That is,

        \begin{axiom}[Archimedian]
            \label{axiom:arch}
            Suppose $x \succ x' \succ x''.$ Then $\exists \alpha, \beta \in (0,1] $ such that $h(x, x'', \alpha) \succ x' \succ h(x, x'', \beta).$
        \end{axiom}

    This is a restriction on how "good" or "bad" a SWP can be. To highlight it's features, it is helpful to think of the extreme case where $\alpha$ is some tiny $\epsilon$ and $\beta$ as being $1-\epsilon$. Then $x' \succ h(x, x'', \beta)$ says that when considering three processes ranked sequentially, the inferior SWP cannot be so bad that mixing a tiny amount of it with the superior SWP makes the combination less preferred than the intermediate process. A similar logic applies to mixing a tiny amount of the superior with the inferior. Viewing this another way, it is as if no matter what process you give me, I can find two other ones "close enough" such that axiom \ref{axiom:arch} holds. The next axiom restricts the impact of mixing on the ranking of processes,

        \begin{axiom}[Independence]
            \label{axiom:independence}
            \( x \succeq x' \implies h(x, x'';\alpha) \succeq h(x', x'';\alpha). \)
        \end{axiom}

    This seems to restrict how diversification can impact the agents preferences. To see this, suppose there are three assets $A, B$ and $C$, whose changes over time are encoded by the stochastic processes $x_A, x_B, x_C \in \mathcal{X}$. Then the axiom says that if you prefer asset A to B then you must also prefer the portfolio of A and C to the portfolio of B and C. Whatever benefit you received from mixing asset C with asset A must be at least as great as the benefit you receive from mixing it with B. Now we can state a helpful lemma.

        \begin{lemma}
            \label{lemma:helpful}
            $\succeq$ on $\mathcal{X}$ satisfies axioms \ref{axiom:consistency} - \ref{axiom:independence} $\iff$
                \begin{enumerate}
                    \item $x \succ x'$ and $0\leq \alpha < \beta \leq 1 \implies h(x, x';\beta) \succ h(x, x';\alpha)$
                    \item $x\succeq x' \succeq x''$ and $x \succ x'' \implies \exists$ unique $\alpha* \in [0,1]$ such that $x' \sim h(x, x'', \alpha*)$ 
                \end{enumerate}
        \end{lemma}
    
        \begin{proof}
            See Appendix \ref{appendix:helpful}.
        \end{proof}

    The previous lemma is critical in proving the following,

        \begin{theorem}[Mixture Space Theorem]
            \label{theorem:MST}
            The complete preorder $\succeq$ on mixture space, $\mathcal{X}$ satisfies axioms \ref{axiom:consistency} and \ref{axiom:independence} $\iff \exists$ linear functional $L:\mathcal{X} \to \mathbb{R}$ such that $x \succeq x' \equiv L(x) \geq L(x').$
        \end{theorem}

        \begin{proof}
            See Appendix \ref{appendix:MST}.
        \end{proof}

    From the proof of Lemma \ref{lemma:helpful} and Theorem \ref{theorem:MST} we find that 
        \begin{equation}
            \label{eq:MST}
            L(x'')=
            \begin{cases}
                \frac{1}{\alpha}, \hspace{0.5cm} x'' \succ x \succ x' \\
                \alpha, \hspace{0.5cm}  x \succ x'' \succ x' \\
                \frac{\alpha}{\alpha - 1}, x \succ x' \succ x''
            \end{cases}  
        \end{equation}
        
    where $\alpha= \sup\{\alpha_0 \in [0,1]: x' \succ h(x, x'';\alpha_0)\}.$ If we can show that $L(x) = E[x]$, then we have a representation similar to that of vNM. However, $E[x] = \int\limits_{supp(\Phi)} x d\Phi$ is not presently well defined. For one, we would need to take the expectation with respect to some point in time. Then we would need to show $L_t(x) = E_t[x]$, and consequently, we would need to define $L_t$. This can be done in the following manner. For each time $t \in \mathcal{T},$ and each $x \in \mathcal{X}$, define the finite stochastic wealth process $x^t:=  \{x_0, x_1, ..., x_t\}.$ Let $\mathcal{X}^t$ denote the collection of all SWPs truncated at time $t$ and suppose $\mathcal{X}^t \subset \mathcal{X}.$ Now, we can retain the consistency axiom but strengthen axioms \ref{axiom:arch} and \ref{axiom:independence} such that they hold for all $t$. Then the mixture space theorem still applies and we could define $L_t$ analogously to equation \ref{eq:MST}. These finite SWPs have associated joint probability distributions $\Phi_t$ and now we could try and show that $L_t(x)=\int\limits_{supp(\Phi_t)}xd\Phi_t$, where, abusing notation, $x$ is taken to be a $t$ period process. But we come to another roadblock. We have made no restrictions on how our SWPs behave, therefore just as in equations \ref{eq:time_rate} and \ref{eq:expected_rate}, neither limit is guaranteed to exist. To overcome this, define $L^2(\Omega):=\{ f: \int\limits_{F_t} f^2(x_t) d\Phi_t < \infty \forall t\}$ to be the set of all Lebesgue square-integrable and monotonic functions f. Now our expectation operator is well defined and if we can show that $L_t \equiv E_t$ then we have a form similar to vNM. Next I impose the following,

        \begin{axiom}[Monotonicity]
            \label{axiom:monotonocity}
            For any \(x, x' \in \mathcal{X} \), if \( x_t(\omega) \geq x_t'(\omega) \forall \omega \in F_t, t \in \mathcal{T}\) then $x \succeq x'.$
        \end{axiom}

    In the context of wealth, this seems rather intuitive and nonrestrictive; I am assuming that an agent would prefer more money to less. Suppose $y:= f(x) = \{f(x_0), f(x_1), ...\}$ and $y':= f(x') = \{f(x'_0), f(x'_1), ...\}$. Then by axiom \ref{axiom:arch}, $y, y' \in \mathcal{X}$ and by monotonicity, $x \succeq x' \iff y \succeq y'.$ Combining the previous axioms, our current setup would be nearly identical to EUT. Our agent would be able to pick from any number of monotonic functions and then left searching for her risk preferences. However, the next postulate ensures this is not the case.

        \begin{axiom}[Ergodicity]
            \label{axiom:ergodic}
            \( x \succ x' \iff \lim\limits_{t\to \infty }\frac{f(x_t(\omega)) - f(x_0)}{t} > \lim\limits_{t \to \infty} \frac{f(x^*_t(\omega)) - f(x_0)}{t} \)
        \end{axiom}

    This axiom states that when ranking two processes, the agent considers how her wealth will grow over the long run. She considers how her wealth is expected to change over time and ranks stochastic wealth processes according to the time average growth rate of some transformation of the SWP. Currently, it may seem that $f$ is arbitrary, and that we have done no better than the expected utility paradigm. But as the next section illustrates, this is not the case. Now we can state the main theorem,

        \begin{theorem}[Ergodic vNM]
            \label{theorem:main}
            $\succeq$ on $\mathcal{X}$ satisfies axioms \ref{axiom:consistency} - \ref{axiom:ergodic} $\iff$
            \begin{enumerate}
                \item \( x \succeq x' \iff E[f(x_t)] \geq E[f(x_t')] \)
                \item \( E[f(x_t)] = \lim\limits_{t \to \infty}\frac{f(x_t) - f(x_0)}{t} \)
            \end{enumerate}
        \end{theorem}

        \begin{proof}
            See Appendix \ref{appendix:main}
        \end{proof}   

    Denoting the transformation as $f$ instead of $u$ is not coincidental. As formulated, the function says nothing about an agent's risk preferences and it is silent as to whether or not the agent prefers early resolution of uncertainty to later. Therefore calling it a utility function does not seem to be in keeping with the spirit of what a utility function encapsulates. Normatively, it's use is easily justified - I assume all agents want to maximize the time average growth of their wealth. Whether or not it is descriptive is an empirical question, but one with a major advantage: changes in the ergodic transformation of wealth are observable, changes in risk aversion are not. 

\section{Ergodic Transformation from SWP}
        In the new set-up, agents begin with a model of how their will will evolve. Knowing that for some finite $t$ their expected growth rate is not necessarily equivalent to the time average growth rate, they seek an appropriate ergodic transformation of the process. If chosen correctly, maximizing the expected value of this process is equivalent to maximizing the time average growth rate of their wealth. To gain some intuition of how one can determine such a transformation in practice, let's consider an example. Suppose changes in wealth follow an arbitrary Ito process,

        \begin{equation}
            dx = a(x)dt + b(x)dW,
        \end{equation}

    where $a$ and $b$ are potentially functions of $x$.\footnote{Full transparency, an additional assumption is now present - I am assuming that the random variables are identically distributed.} Then our problem is to find a transformation $f$, such that
    
        \begin{equation}
            df(x) = \alpha dt + \beta dW,
        \end{equation}

    where $\alpha$ and $\beta$ are constants defined as
    
        \begin{equation}
            \label{equation:alpha}
            \alpha := a(x) \frac{\partial f}{\partial x} + \frac{b^2(x)}{2}\frac{\partial ^2 f}{\partial x^2}
        \end{equation}
    and

        \begin{equation}
            \label{equation:beta}
            \beta := b(x)\frac{\partial f}{\partial x}.
        \end{equation}

    That is, our transformed stochastic process is a Levy process (stationary and independent increments). By application of Ito's Lemma, \citet{petersEvaluatingGamblesUsing2016} showed that for a general dynamic to have an associated ergodic transformation, 

        \begin{equation}
            \label{equation:condition}
            a(x) = \frac{\alpha}{\beta}b(x) + \frac{b(x)b'(x)}{2}
        \end{equation}

    As a concrete (but intentionally contrived) example, suppose the stochastic wealth process is governed by the dynamic

        \begin{equation}
            dx = x^{\gamma}[1 + \frac{\gamma}{2}x^{\gamma - 1}dt + dW]
        \end{equation}

    To economize notation, I will informally write $f'$ for $\frac{\partial f}{\partial x}.$ Then 

        \begin{equation}
            a(x) = x^{\gamma} \frac{\gamma}{2}x^{2\gamma - 1}
        \end{equation}
    
    and
    
        \begin{equation}
            b(x)=x^{\gamma }.
        \end{equation}
    
    Substituting these expressions into equation \ref{equation:condition} yields,
        \begin{equation}
            b(x) + \frac{1}{2}b(x)b'(x) 
        \end{equation}

        \begin{equation}
            =x^{\gamma} + \frac{\gamma}{2}x^{2\gamma -1}
        \end{equation}

        \begin{equation}
            = a(x),
        \end{equation}

    therefore there exists an ergodic transformation. Then substituting the expression for $b(x)$ from equation \ref{equation:beta} into equation \ref{equation:alpha} yields a differential equation,

        \begin{equation}
            \label{eq:diff_eq}
            f' = \frac{1}{b(x)}
        \end{equation}
    
        \begin{equation}
            = x^{-\gamma}
        \end{equation}
    
        \begin{equation}
            \iff f(x) = \frac{x^{1-\gamma}}{1-\gamma} + C , \hspace{0.5cm} \gamma \neq 1
        \end{equation}

    This should be a very familiar representation and highlights why I have chosen $f$ instead of $u$ to represent the transformation. In economics, an agent with a utility function $u(x)= \frac{x^{1-\gamma}}{1-\gamma}$ is said to be risk averse. More specifically, they are said to exhibit constant relative risk aversion, the level of which is captured by the size of $\gamma.$ However, here it does not seem fair to call an agent risk averse. They are neither optimistic nor pessimistic about the state of nature tomorrow, they simply want to maximize the growth rate of their wealth.

\section{Revisiting Discounted Utility}
    The following section loosely describes how once we have an ergodic transformation, we can revive a formulation mathematically equivalent to discounted utility. Consider a finite but continuous time model where $t \in [0, T].$ Let $V(f(x_{t_0 + \Delta t}))$ denote the value of an asset after time $\Delta t$ has elapsed. Assume that the function $f$ is the transformation that renders the the dynamic wealth process (the asset) ergodic. Now consider the value at some point $s>t$, i.e. $V(f(x_{t_0 + \Delta s})).$ Since the transformed process is ergodic, it has independent increments and $V(f(x_{t_0 + \Delta s}))$ only depends on $\Delta s$. Then $\frac{V(f(x_{t_0 + \Delta s}))}{V(f(x_{t_0 + \Delta t}))}$ doesn't depend on $t_0$. Now, letting $t_2 = t_1 + \delta,$ we seek a measurable function $\Psi$ satisfying 

        \begin{equation}
            \frac{\Psi(f(x_{t_1}))}{\Psi(f(x_{t_0}))} = \frac{\Psi(f(x_{t_{1 + \delta}}))}{\Psi(f(x_{t_{\delta}}))}
        \end{equation}

        \begin{equation}
            \iff \Psi(f(x_{t_{\delta}}))\Psi(f(x_{t_{1}})) = \Psi(f(x_{t_{1 + \delta}}))\Psi(f(x_{t_{0}}))
        \end{equation}

        \begin{equation}
            \iff \Psi(f(x_{t_{2-1}}))\Psi(f(x_{t_1})) = \Psi(f(x_{t_{1 + \delta}}))\Psi(f(x_{t_0}))
        \end{equation}

    Then letting $\Psi(f(x_{t_0}):=\beta$,

    \begin{equation}
        \Psi(f(x_{t_1})) = \beta \frac{\Psi(f(x_{t_{1 + \delta}}))}{\Psi(f(x_{t_{\delta}}))}.
    \end{equation}

    This is a Cauchy exponential functional equation, having the well known solution, 

        \begin{equation}
            \Psi(f(x_{t_0 + \Delta t})) = \beta e^{-\alpha \Delta t}
        \end{equation}

    for some constant $\alpha$, which depends on the function $f$. Generalizing, we have an equation for the value of an asset after time $\Delta t$ has passed,

        \begin{equation}
            V(f(x_{t_0 + \Delta t})) = \beta e^{-\alpha \Delta t}.
        \end{equation}

\section{Conclusion}

    The new framework is not without criticism. For example, we are taking the limit as $t \to \infty$, but life is pre-asymptotic. Additionally, have left out any notion that an agent might have preferences for early resolution of uncertainty. One way to proceed is given two dynamics with the same time average growth rate, follow \citet{petersOptimalLeverageNonergodicity2011} and determine the characteristic time scale of each. Then we could suppose an agent prefers the SWP with the shorter time to convergence. In the preceding arguments, I have assumed that the agent knows perfectly well the dynamics of the stochastic wealth process. Interpreting this in the form of vNM, we would say that uncertainty is purely objective. However, it might be the case that agents are concerned about model misspecification. Such a scenario is precisely what is considered in \citet{hansenRobustControlModel2001} and future work should consider extending the current framework in the direction of ambiguity aversion and robust preferences. Notwithstanding, the new paradigm does not leave any room for interpreting $\gamma$ as a coefficient of risk aversion. Applied economists might lament that they have lost a degree of freedom, but this can be rectified. Suppose the agent has determined that the appropriate ergodic transformation is $f(x)=\frac{x^{1-\gamma}}{1-\gamma}.$  Suppose $\lim\limits_{T \to \infty}E_T[f(x)] = c$ and there is a constant process $x^c$ where your wealth grows by $\frac{c}{T}$ each period. Then an agent is risk averse if $x^c \succeq x$ and one agent is more risk averse than another if they would accept a smaller amount, say $c'$. Their utility functions then would be something like $u(x)=\frac{x^{1-\lambda \gamma}}{1 - \lambda \gamma}.$  However, I want to emphasize that this is if and only if $f(x)=\frac{x^{1-\gamma}}{1-\gamma}$. If the ergodic transformation was something else, say $f(x)=1-\exp^{-\alpha x}$ then the utility function would need to be redefined analogously. 

\newpage
\printbibliography
\newpage

\appendix
\appendixpage
    \section{Copehnagen Experiment}
    \label{appendix:CE}
        To describe their experiment, let us adopt the following definitions. A \textit{gamble} is a pair of different images. In the Copenhagen Experiment (CE), there are 18 distinct images, each corresponding to a unique change in wealth. A \textit{trial} is a pair of gambles. A \textit{game} is characterized as additive or multiplicative and consists of a passive and active phase. In the Copenhagen Experiment, each subject played two games - one multiplicative and one additive. During the passive phase, the agent is shown a sequence of 9 images 37 times. As each image appears on the screen, the agent observes her wealth increase or decrease. The experiment administrators stress to the subject the importance of learning the relationship between the images and changes in wealth. During the active phase, the agent participates in 312 \textit{trials} where he or she is shown two unique gambles and chooses one or the other. With equal probability, one of the images comprising the selected gamble will be assigned to her, however, she does not know which one. Afterwards, 10 of the assigned images are randomly selected and their corresponding wealth changes are applied to the subjects endowment to calculate terminal wealth.

        \begin{figure}[hb]
            \centering
            \includegraphics[width=0.3\textwidth]{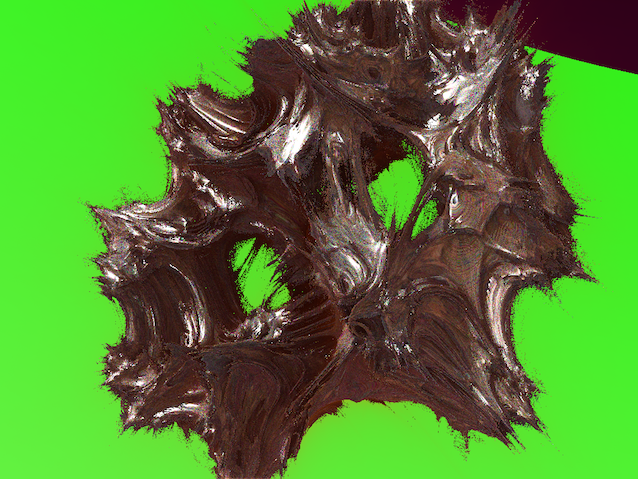}
            \caption{Example of 'additive' fractal}
            \label{fig:add1}
        \end{figure}

    \section{Comparison Models}
    \label{appendix:comp_models}
    Taking the advice of \citet{doctorEconomistsViewsErgodicity2020}, in this section I will compare the setting of the Copenhagen Experiment (CE) to those of classical utility models. The purpose is to elucidate the workflow of decision making in these models and highlight the additional assumptions/restriction needed to map CE into existing theories. Throughout, I will assume there exists only a single consumption good and that the agent has a known initial endowment. I assume that the problem the agent faces is one of selecting an optimal strategy, before the resolution of any uncertainty. The first model I consider, vNM in a single period setting, is obviously not suitable to dynamic or intertemporal decisions, but lays the foundation for future comparisons. 

    \subsection{Static vNM}
        \subsubsection{Description}
            The agent must take some action today that impacts $\tilde{c_T}$, her consumption tomorrow, a random variable. Suppose that the agent has primitive preferences over consumption \textit{levels}, $c_T \in C \subseteq \mathbb{R},$ and the preference relation $\preceq$ on $C$ satisfies
            
                \begin{axiom}[Consistency]
                    \(\preceq\) is a complete preorder on \(C\).
                \end{axiom}
                
                \begin{axiom}[Continuity]
                    The sets \( \{c' \in C | c \preceq c' \} \) and \( \{c' \in C | c' \preceq c \in C\} \) are closed.
                \end{axiom}
            
                \begin{axiom}[Independence]
                    Define \( h(c, c'):= \alpha c + (1-\alpha)c', \alpha \in (0,1) \). Then \( c \preceq c' \implies h(c, c'') \preceq h(c', c''). \)
                \end{axiom}
            
            therefore the function $u$ cardinally represents the relation $\preceq$. However, since consumption tomorrow is a random variable, the agent considers utility of final consumption with respect to each lottery that would be induced by her action today, and selects the one yielding the maximum utility of consumption in expectation. This ranking of actions induces a preference relation over lotteries,
            
                \begin{equation}
                    \label{eq:vNM}
                    l \succeq' l' \iff U(l):= E_l[u(c)] \geq E_{l'}[u(c)] =: U(l'),
                \end{equation}
            
            where $\succeq'$ again obeys the axioms above (suitably adapted). Then the agents decision at $t=0$ is the one which induces the most preferred lottery. Substituting in the constraint that final consumption cannot exceed terminal wealth and assuming that $u$ is strictly increasing and concave, equation \ref{eq:vNM} indirectly defines a preference relation over terminal wealth levels. Note, this final assumption is often attached to notion of risk aversion. An agent is defined to be risk averse if and only if $u$ is strictly increasing and concave.
            
        \subsubsection{Adaptation}
            Adapting this method to CE will require a number of simplifications. Suppose the subjects only make a single decision at the very beginning and then their endowment increases/decreases by a stochastic amount to arrive at terminal wealth levels. We would need to assume the agent makes a single choice between gambles at initiation, which then leads to a sequence of stochastic changes in wealth, which we need to collapse into a single terminal change. One possibility is that the agent's choice between the left or right gamble is seen as triggering the spin of roulette wheel one or roulette wheel two. If we imagine that an agent agreed to simply flip a coin at each decision, we could obtain the induced lotteries associated with choosing left or right at initiation and then flipping a coin the remaining number of trials. A shortcoming of this model that will be repeated throughout is that the agent does not know the exact form of $u$. The axioms of vNM, while seemingly too stringent, do not have enough bite to yield a particular function. 

            However, \citet{krepsNotesTheoryChoice2018} provides a heuristic simplification. If the agent can answer in the affirmative to the following three questions: 
            
                \begin{enumerate}
                    \item Do I like more money to less?
                    \item Am I risk averse to changes in consumption along some range? 
                    \item Do I have slight decreasing absolute risk aversion over this same range?
                \end{enumerate}
            
            then their utility function belongs to a class of exponential functions parameterized by $\lambda$, their level of risk aversion. That is, $u(z):= -e^{-\lambda z},$ and their level of risk aversion can be obtained via repeated introspection. For example, first find your certainty equivalent for a simple gamble and then repeat this process over and over, each time obtaining closer approximations to your level of risk aversion.

    \subsection{Payoff Vector Approach}

        \subsubsection{Description} 
            This method supposes an agent has primitive preferences over deterministic payoff \textit{vectors} $\textbf{z} \in \mathbb{R}^T$, where each $z_t$ denotes a deterministic payoff at time $t$. Her preferences obey the classical axioms of vNM therefore they are representable by a cardinal utility function $u:\mathbb{R}^T \to \mathbb{R}$. In a dynamic decision problem, the agent chooses a strategy, or sequence of actions, which restricts the space of potential payoff vectors. In other words, any strategy induces a joint probability distribution over the space of payoff vectors. Then she ranks each strategy according to its corresponding expected utility of consumption given the induced measure, $\mu$. That is she considers, 
            
            \begin{equation}
                U(\textbf{z}, s):= \int\limits_{supp(\mu)}u(\textbf{z})d\mu
            \end{equation}

            where $supp(\mu)$ is the set of payoff vectors possible given the sequence of actions specified by strategy $s$. 

        \subsubsection{Adaptation}
            Adapting this to the CE requires us to assume that each agent has preferences over payoff vectors satisfying vNM. Then at each point in time the agent's decision is to select the left gamble or the right gamble. Culling these selections together would represent a single strategy which then induces a probability distribution over payoff vectors. Our agent would then select the strategy yielding the most favorable outcome in expectation. However, there are three problems. (1) In the CE, the agent does not know how many trials she will play. This is important because in general, the joint distribution of an N-period stochastic process is not equivalent to that of an N+1 period process. But suppose we altered the experiment so that the decision maker knew the number of trials. (2) The uncertainty for each gamble does not resolve until $t=T$, whereas the payoff vector approach assumes that uncertainty resolves in the same period. We could augment the experiment once more and ask what if they subject learned their outcome after each selection? In this case the agent, normatively at least, would be able to solve for the optimal strategy via backward induction. Descriptively however, the complexity of the branching process in the Copenhagen Experiment makes backward induction impossible, even for most supercomputers. (4) And finally, the agent would know their exists a utility function representing her preferences, but she is left wondering what the form of this function is, and if found, then what is her level of risk aversion? The simplification provided above does not carry over as neatly as before, however it would still be theoretically possible to approximate ones level of risk aversion.
                
            There are additional well-known problems with this approach that apply more broadly. The works of \citet{mossinNoteUncertaintyPreferences1969}, \citet{drezeConsumptionDecisionsUncertainty1972a} and \citet{spenceEffectTimingConsumption1972} (and many others) pointed out two stylized facts about atemporal expected utility models. (1) They obscure the importance of timing of resolution of uncertainty and (2) the induced preference relation does not satisfy the axioms of vNM. As for (2), note that as was the case with the single-period model, we could substitute in final wealth as the sum of the payoffs to arrive at an implied preference relation over lotteries of terminal wealth. However, this implied preference relation does not in general satisfy the independence axiom of vNM, as cogently pointed out in \citet{mossinNoteUncertaintyPreferences1969}. Additionally, \citet{krepsTemporalNeumannmorgensternInduced1979} gives an example of why it is doubly wrong to use atemporal vNM in \textit{any} dynamic model, which led them to formulate the following model of decision making.

    \subsection{Temporal vonNeumann-Mergernstern}

        \subsubsection{Description}
            \label{section:temp_vNM}
            \citet{doctorEconomistsViewsErgodicity2020} posit that the Copenhagen Experiment and EE should be compared (normatively) to the predictions of temporal vNM. To describe this model we will need to adopt a bit of notation. Denote the set of possible payoffs at time $t \in [0, T]$ as $Z_t$, a complete separable metric space. Beginning at time $T$, denote the set of Borel probability measures on $Z_T$ as $m(Z_T)$. Define $D_T := m(Z_T)$ to be the set of actions at time $T$ and endow this set with the Prokhorov metric (i.e. the metric of weak convergence). Note, this set is a mixture space. Then take this set and generate $X_T$, the class of all non-empty closed subsets of $D_T$ and endow it with the Hausdorff metric. Then $\forall t < T$  recursively define $D_t := m(Z_{t-1} \times X_t).$ Suppose we are standing at $t=1$. Then we can think of $X_{t+1}$ as the $\sigma$-ring generated by all one-step-ahead probability measures induced by our previous actions. Our agent keeps track of where he is at in time by referencing her realized payoff history, $y_t:=\{z_0, z_1, ... , z_{t-1}\} $. I denote the collection of potential histories up to time $t$ as $Y_t.$ Think about it as if the agent is able to pause time at $t-\epsilon$, look at the payoffs he has received, plus consider potential payoffs that he is about to receive. Then fixing $t$, the authors consider $\preceq_{y_t},$ an agent's preferences conditional on realized payoffs and possible current payoffs. In this sense, $y_t$ can be thought of as representing the information known by the agent. 
            
            The authors assume that agents have primitive preferences over temporal lotteries of payoffs and impose the axioms of vNM on each temporal preference relation. For any $t$ and $y_t$,

            \begin{axiom}[Static Consistency]
                \label{axiom:static_consistency}
                \(\preceq_{y_t}\) is a complete preorder on \(D_t\).
            \end{axiom}

            \begin{axiom}[Continuity]
                \label{axiom:continuity}
                The sets \( \{d_t' \in D_t | d_t \preceq_{y_t} d'_t \} \) and \( \{d_t' \in D_t | d'_t \preceq_{y_t} d_t \} \) are closed in the weak topology.
            \end{axiom}

            \begin{axiom}[Independence]
                \label{axiom:substitution}
                Define \( h(d, d'):= \alpha d + (1-\alpha)d', \alpha \in (0,1) \). Then \( d \preceq_{y_t} d' \implies h(d, d'') \preceq_{y_t} h(d', d''). \)
            \end{axiom}

            Notice the subscript on the preference relation. If each \textit{temporal} preference relation obeys the above axioms, then each can be represented by an expected utility formulation. Then to knit together these preference relations across time they impose the following:

            \begin{axiom}[Temporal Consistency]
                \label{axiom:temp_consistency}
                \( \forall t, y \in Y_t, z \in Z \) and \( x, x' \in X_{t+1} \), \( (z, x) \succeq_y (z, x') \) at time \( t \iff x \succeq_{y,z} x' \) at time \( t+1 \) 
            \end{axiom}
        
            With preferences satisfying the above axioms, the dynamic choice problem can be solved by backward induction. Essentially, the agent would compute their utility for each terminal node and then their expected utility at all penultimate nodes. From this calculation they would select the time $T-1$ action that induces the maximum expected utility at each possible state. Then treating the time $T-1$ nodes as the terminal nodes they would repeat the process until they arrive at $t=0$, at which point they will have collected the conditionally optimal decisions at each point in time. Culling these together, the agent obtains the set of all optimal strategies for any contingency.  

            In figure *** below, I have reproduced the example from \citet{krepsTemporalNeumannmorgensternInduced1979}, highlighting the optimal strategy. As a comparison, figure *** depicts the Copenhagen Experiment in the language of \citet{krepsTemporalResolutionUncertainty1978}.

            [Insert Figure *** here.]

        \subsubsection{Adaptation}
            As evident from figure ***, if we again assumed that the agent's know how many decisions they will be making, the CE fits nicely into this framework. We would also need to assume that agents prefer early resolution to later, but this does not seem prohibitive. Normatively, this framework is sure to produce the optimal strategy. Objections on descriptive grounds remain, as well as the usual objection that an agent is left searching for a particular utility function and for her level of risk aversion. It bears emphasis that the primary motivation for \citet{krepsTemporalResolutionUncertainty1978} was the notion that agents typically prefer earlier resolution of uncertainty to later and that this possibility is not captured by the payoff vector approach. That induced preferences did not satisfy independence was an ancillary consideration and one that their axiomatization explicity precludes. Roughly, the required condition was that preferences be time separable and therefore the agent's utility function have the form $U(c_0, c_1)=f(c_0) + g(c_0)h(c_0 + c_1).$ This is in spite of the fact that \citet{mossinNoteUncertaintyPreferences1969} provides an example of how it can be perfectly rational for ones induced preferences over wealth distributions to not have an expected utility representation. 

    \subsection{Conclusion}
            Note that in the example from section \ref{section:temp_vNM}, we are assuming without justification that the agent has the utility function, $U(z_0, z_1):=\sqrt{z_0 + z_1}.$ The hope then is that an axiomatization of EE will give conditions that prescribe a specific functional form, i.e., pin down exactly when an agent should have square root utility, log utility, etc. This entails a significant loss of generality. Indeed, some may object on the grounds that there will appear to no longer be any room for idiosyncratic risk aversion, while countless experiments have reported strong evidence that this is an important factor in decision making. This is an objection that can be remedied, but only after we have determined exactly what is required.

    \section{Proof of Lemma \ref{lemma:helpful}}
    \label{appendix:helpful}
        All we need to verify is that $h(x, h(x', x'';\beta);\alpha) = h(h(x, x';\frac{\alpha}{\alpha + \beta - \alpha \beta},x'')),$ which should be easy to see since our random variables are mapping to real numbers and addition over the reals is associative. Once verified, the proof of the lemma is nearly identical to that found on p.47 of \citet{krepsNotesTheoryChoice2018}. Part (1) of the lemma implies that if $\alpha^*$ exists then it is unique. Part (2) is a proof by construction where ultimately we find that $\alpha^*= \sup\{\alpha \in [0,1]: x' \succ h(x, x'';\alpha)\}.$

    \section{Proof of Theorem \ref{theorem:MST}}
    \label{appendix:MST}
        With Lemma \ref{lemma:helpful} in hand, the proof again follows closely to that outlined in \citet{krepsNotesTheoryChoice2018}. I only provide a sketch.
            \begin{enumerate}
                \item Non-degenerate $\implies \exists x, x' \in \mathcal{X}$ such that $x \succ x'.$
                \item Define $L(x):=1, L(x'):=0.$
                \item Then for any other $x'' \in \mathcal{X}$ we have three cases to consider:
                    \begin{enumerate}
                        \item $x'' \succ x \succ x' .$ Then by Lemma \ref{lemma:helpful}, $\exists$ unique $\alpha$ such that $h(x'', x;\alpha) \sim x.$ So define $L(x''):= \frac{1}{\alpha}$
                        \item $ x \succ x' \succ x''$. Then by Lemma \ref{lemma:helpful}, $\exists$ unique $\alpha$ such that $h(x'', x) \sim x'.$ So define $L(x''):=\frac{\alpha}{\alpha - 1}$
                        \item $x \succ x'' \succ x'$. Again,  by Lemma \ref{lemma:helpful}, $\exists$ unique $\alpha$ such that $h(x, x';\alpha) \sim x''.$ Therefore, define $L(x''):=\alpha.$
                    \end{enumerate}
            \end{enumerate} 

    \section{Proof of Theorem \ref{theorem:main}}
    \label{appendix:main}
        I have only worked out a sketch of the proof, which I provide.
        \begin{proof}           
            First, fix t and recall that extending vNM to the space of non-simple distributions requires a suitable topology. This is not an innocuous decision, as the topology will dictate continuity and convergence - the harder it is to converge, the weaker our continuity axiom becomes. However, here we have defined preferences on SWPs and by monotonicity have restricted the relation to $L^2(\Omega)$, which is already a topological vector space (Hilbert space). Then by axioms \ref{axiom:consistency} - \ref{axiom:monotonocity}, $\exists f \in L^2(\Omega)$ such that $x \succ x' \iff L(f(x_t)) = \alpha = E[f(x_t)] > E[f(x'_t)] = \alpha' = L(f(x'_t))$ (this is almost equivalent to vNM). But by the EA, $x > x' \iff \lim\limits_{t \to \infty}\frac{f(x_t) f(x_0) }{t} := g > g' =: \lim\limits_{t \to \infty}\frac{f(x'_t) - f(x_0)}{t}$ Therefore $g = \alpha \implies E[f(x_t)] = \lim\limits_{t \to \infty}\frac{f(x_t) - f(x_0)}{t}.$ Thus, $f$ is an ergodic transformation of $x$. 
            
            Axiom \ref{axiom:ergodic} begs for comparison to the dynamic consistency axiom of \citet{krepsTemporalResolutionUncertainty1978}. There, although the axioms are able to equate the linear functional to the expectation operator, the form of $u$ remains a free parameter. The representation holds for any measurable $u$. Here, axiom \ref{axiom:ergodic} ensures that $f$ is the ergodic transformation, i.e. the one that pulls out the growth rate from the dynamic process. 
        \end{proof}

        As an aside, the following definition of an ergodic transformation will be useful in formalizing the proof. Let $(\Omega, \mathcal{A}, \mu)$ and $(\Omega', \mathcal{A}', \mu')$ be two probability spaces. Then I will say the transformation $T:\Omega \to \Omega'$ is measurable if it is invertible and if
    
                \begin{equation}
                    T^{-1}(\mathcal{A}') \subset \mathcal{A}.
                \end{equation}
    
            I will call a transformation measure preserving if it is measurable and if
    
                \begin{equation}
                    \mu(T^{-1}(A'))=\mu'(A') \forall A' \in \mathcal{A}'.
                \end{equation}
    
            And finally, I define an ergodic transformation as any measure preserving morphism $T:\Omega \to \Omega$ satisfying 
    
                \begin{equation}
                    T^{-1}A = A \iff \mu(A) = 0 \hspace{0.25cm} \text{or} \hspace{0.25cm} 1,              
                \end{equation}
    
            for any $A \in \mathcal{A}$. In words, $T$ is an ergodic transformation if the only corresponding inverse-invariant sets are those assigned measures 0 or 1. Comparing this formulation to that of equation (\ref{equation:ergodicity}) highlights how the property of ergodicity is dependent on the measure, the Borel $\sigma$-algebra, and the state space. As a concrete example, the identity transformation is obviously measure preserving. It is then ergodic if and only if every Borel-measurable subset is assigned a measure of zero or one. This is quite a strong requirement and foreshadows complications that will arise when considering joint distributions associated with a sequence of random variables.

\end{document}